\documentclass[aps,prb,twocolumn,amsmath,amssymb,floatfix,footinbib,bibnotes,groupedaddress,longbibliography]{revtex4-1}

\usepackage{soul}
\usepackage{bm}
\usepackage{epsf}
\usepackage{amssymb}
\usepackage{esint}
\usepackage{amsmath}
\usepackage{comment}
\usepackage{graphicx}
\usepackage{rotating}
\usepackage{epsfig}
\usepackage{psfrag}
\usepackage{amsmath}
\usepackage{subfigure}
\usepackage{nicefrac}
\usepackage{bm}
\usepackage[usenames,dvipsnames]{color}
\usepackage{soul}
\usepackage{multirow}
\usepackage{titlesec}
\usepackage{tabularx}

\usepackage[colorlinks,citecolor=blue,urlcolor=blue,linkcolor=blue]{hyperref}

\def \up{\uparrow}
\def \down{\downarrow}

\global \def \aa#1 {\begin{align} #1 \end{align}}

\newcommand {\apgt} {\ {\raise-.5ex\hbox{$\buildrel>\over\sim$}}\ }
\newcommand {\aplt} {\ {\raise-.5ex\hbox{$\buildrel<\over\sim$}}\ }

\begin{document}

\title{Quantum criticality and confinement in weak Mott insulators}

\author{Eyal Leviatan}
\affiliation{Department of Condensed Matter Physics, Weizmann Institute of Science, Rehovot 7610001, Israel}

\author{David F. Mross}
\affiliation{Department of Condensed Matter Physics, Weizmann Institute of Science, Rehovot 7610001, Israel}

\begin{abstract}
Electrons undergoing a Mott transition may shed their charge but persist as neutral excitations of a quantum spin liquid (QSL). We introduce concrete two-dimensional models exhibiting this exotic behavior as they transition from superconducting or topological phases into fully charge-localized insulators. We study these Mott transitions and the confinement of neutral fermions at a second transition into a symmetry-broken phase. In the process, we also derive coupled-wire parent Hamiltonians for a non-Abelian QSL and a $\mathbb{Z}_4$ QSL.
\end{abstract}

\maketitle

\textit{Introduction}---The Mott transition~\cite{Imada1998} of spinful fermions is central to numerous compelling phenomena in quantum many-body physics. High-temperature superconductivity in the cuprates, for example, arises near the transition between a non-magnetic metal and an antiferromagnetic insulator.\cite{cupratereview1} In addition, ‘weak Mott insulators,’ i.e., systems just barely on the insulating side of the quantum phase transition (QPT), provide fertile ground for exotic forms of quantum magnetism, as observed in various organic compounds.\cite{organicsreview,Furukawa2015} A small charge gap promotes multi-spin interactions, stabilizing QSL ground states with fractional ‘spinon’ excitations.\cite{qslbalents,qslsavary,qslwen,qslzhou,qslknolle,qslbroholm} 

Many properties of QSLs, such as deconfined spin-$\frac{1}{2}$ excitations, are naturally present in weakly interacting metals or superconductors. There, they are the electronic or Bogoliubov quasiparticles, respectively. In the weak Mott insulator, spinon excitations may fruitfully be viewed as an inheritance of the nearby itinerant phase. Within this framework, QSLs arise when electrons discard their charge but evolve otherwise smoothly across the QPT. Similarly, a singlet Cooper pair relates to a dimer (valence bond) in a spin model. The natural fate of a superconductor undergoing a Mott transition is thus either a valence bond solid (VBS) with frozen dimers or a QSL with fluctuating dimers.\cite{Anderson1973}

We focus on two-dimensional systems of spin-$\frac{1}{2}$ fermions or bosons at an average filling of one particle per unit cell. When such systems undergo a Mott transition, one of two things must happen concomitantly with the localization of unit charge to each site: either some symmetry breaks spontaneously or a QSL forms. Experimentally, Mott transitions are typically first order. Theoretical studies of these QPTs are challenging due to the lack of any small parameter---the transition occurs when interaction strength and bandwidth are comparable. Still, field-theoretical analyses have shown that second-order transitions are also possible in both cases.\cite{Senthil2000,Lannert2001,Balents2005a,Balents2005b,senthil2008}

\begin{figure}[t]
\centering
 \includegraphics[width=\columnwidth]{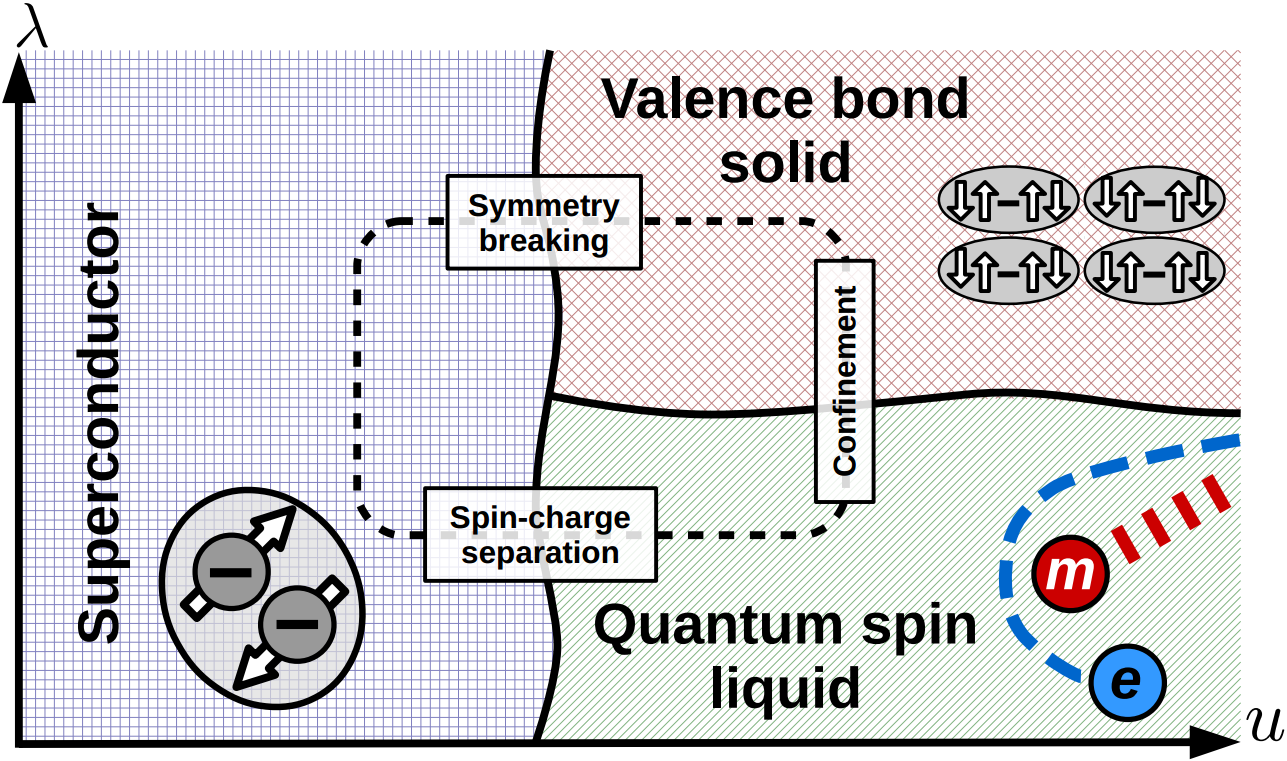}
 \caption{Phase diagram of a superconductor that undergoes a Mott transition upon increasing the on-site repulsion $u$. The parameter $\lambda$ modifies the spin correlations within the superconductor. Depending on its value, the insulating phase may be a valence bond solid or a $\mathbb{Z}_2$ spin liquid. The coupled-wire formalism affords us theoretical control along the dashed line, passing through all phases and phase transitions. \label{fig:phase_diagram}}
\end{figure}

A prototypical phase diagram of electrons undergoing a Mott transition is illustrated in Fig.~\ref{fig:phase_diagram} for the example of a superconductor. When both sides of the QPT are conventional phases, it is either first-order or exhibits `deconfined' criticality.\cite{Senthil2004} By contrast, the Mott transition into a QSL may be governed by the critical point of the classical 3D XY model~\footnote{The universality class is known as XY$^*$, but the critical theory is the same as for XY.\cite{Senthil2000,Grover2010,Isakov2012}} Finally, QSLs and topologically-trivial Mott insulators are separated by a confinement transition. 

To describe the Mott transition, we begin deep within a superconductor or topological insulator. We locally deform its Hamiltonian to write it as a sum of two parts, $H^\text{charge}$ and $H^\text{spin}$, which commute up to irrelevant contributions. 
The `spin' part does not involve charge transfer between different sites. Consequently, the system remains in its ground state as charge carriers localize due to strong on-site repulsion. The competition between the latter and $H^\text{charge}$, projected onto the ground state of $H^\text{spin}$, then characterizes the Mott transition. In one dimension, this property is quite generic; Luttinger liquids factorize into charge and spin sectors, and phase transitions in the former do not affect the latter. 

To carry out these steps in two dimensions we build on the coupled-wire framework.\cite{Kane2002} In particular, the Mott transition and subsequent confinement transitions are accessible in a well-controlled way. Known properties of QPTs out of Abelian and non-Abelian QSLs are reproduced in an almost pedestrian manner, without reference to gauge theories.\cite{Xu2011} Additionally, we derive the critical properties of two QPTs that were not previously discussed: (i) between s-wave superconductor and VBS (ii) between bosonic Laughlin state and chiral QSL. Crucially, for each QPT, we construct a physically sensible microscopic model, i.e., one that is local and only involves two-body interactions. The construction of such models, which may be tested numerically, is beyond all existing approaches to exotic phases near the Mott transition.\\

\textit{Superconductor and VBS---}The conceptually clearest example is the Mott transition out of an s-wave superconductor. To describe it, we begin with an array of one-dimensional quantum wires in the Luttinger liquid phase. At low energies, spin-$\sigma$ electrons near the right (R) or left (L) Fermi points are annihilated by $\psi_{y,\sigma,R/L}$. The singlet Cooper-pair operator is $\hat\Delta_y = \psi_{y,\up,R}\psi_{y,\down,L}-\psi_{y,\down,R}\psi_{y,\up,L}$. We couple neighboring wires via pair-hopping, i.e., $H_\text{SC} = g \hat \Delta_{y+1}^\dagger\hat \Delta_{y} + \text{H.c.}$ (Here and throughout, we lighten the notation by leaving the summation over wires and integration along the wire implicit. The wire index $y$ will also be suppressed unless needed.) When $H_{\text{SC}}$ is relevant in the renormalization group sense or has a large coefficient, a superconductor with $\langle \hat \Delta \rangle \neq 0$ and a hard spin gap arises.

Two-particle intra-wire umklapp processes described by $H_{\text{Mott}}=u \psi_{\up,R}^\dagger\psi_{\down,R}^\dagger\psi_{\up,L}\psi_{\down,L}+\text{H.c.}$ induce the Mott transition. Microscopically, they arise from density-density interactions between electrons, e.g., on-site repulsion in a Hubbard model. At half-filling, repulsion favors localizing the electrons on each wire and thus competes against pair-hopping. When the former prevails, it suppresses the latter, but the spin gap may persist. Indeed, at second order in $g$, $H_\text{SC}$ generates $H^\text{spin}_\text{SC}\sim \hat \Delta_{y}^\dagger\hat \Delta_{y} = \psi^\dagger_{\up,R}\psi_{\up,L}\psi^\dagger_{\down,L}\psi_{\down,R} + \text{H.c.}$ This intra-wire interaction opens a spin gap in Luttinger liquids and survives the Mott transition by not transferring charge.

To characterize the insulator and the QPT, we employ Abelian bosonization.\cite{GiamarchiBook2004,GogolinBook2004} The charge and spin degrees of freedom are encoded by canonically conjugate long-wavelength operators $\theta_{y}^{c/s}$ and $\varphi^{c/s}_{y}$; we use the convention where their densities are $\rho_{c/s} = \frac{1}{\pi}\nabla\theta^{c/s}$. The Cooper-pair operator is, then, $\hat \Delta = e^{i \varphi^c}\cos [2\theta^s]$. 
The intra-wire interactions introduced above are $H_{\text{Mott}} \sim \cos[4\theta^{c}]$ and $H^\text{spin}_\text{SC}\sim\cos[4\theta^s]$. According to these two terms alone, each wire breaks translation symmetry independently, and the ground state is macroscopically degenerate. Residual inter-wire couplings, the leading of which is $\delta H_{\text{VBS}}^\text{spin} =u' \cos[2\theta^c_{y+1}+2\theta^c_{y}]\cos[2\theta^s_{y+1}+2\theta^s_{y}]$, will lock the order parameters of individual wires into a global symmetry-breaking pattern. The resulting VBS ground state does not exhibit topological order, i.e., fractional excitations are confined.

By construction, $H^\text{spin}_\text{SC}$ does not experience competition from $H_\text{Mott},\delta H_{\text{VBS}}^\text{spin},$ or $H_\text{SC}$, i.e., the latter is the `charge' Hamiltonian. The transition is thus governed by $H_{\text{SC--VBS}} \equiv \langle H_\text{Mott}+\delta H_{\text{VBS}}^\text{spin}+H_\text{SC}\rangle_{H^\text{spin}_\text{SC}}$. Here, projection onto the `spin' ground state amounts to replacing all instances of the pinned operators $\theta^s$ by $c$-numbers; we find
\begin{align}
\label{eq.SC_VBS}
 \nonumber H_{\text{SC--VBS}} = & u'\cos[2\theta^c_y+2\theta^c_{y+1}]+u\cos[4\theta^c_y]\\
 & +\cos[\varphi^c_{y+1}-\varphi^c_{y}]~.
\end{align}
This Hamiltonian also describes the transition between an easy-plane antiferromagnet and a spatially anisotropic VBS.\cite{Senthil2004} Its ultimate fate is believed to be a first-order transition.\cite{Kuklov2006,Kragset2006,Chen2009,DEmidio2016,DEmidio2017} Explicitly breaking the wire translations permits the lower-order term $\cos[2\theta^c_y]$, placing the transition into the universality class of the 3D XY model.\\

\textit{Superconductor and $\mathbb{Z}_2$ QSL---}We now modify the parent superconductor to obtain a Mott transition into a deconfined phase. As we will see, a QSL arises from a phase with the order parameter $\hat{\Delta}'_{y} = \psi_{y,\uparrow,R} \psi_{y-1,\downarrow,L}$ for odd $y$ and with $R \leftrightarrow L$ for even $y$. Pairing is induced by Cooper-pair hopping, i.e., $H_{\Delta'}=g_y\hat \Delta_{y+1}^{\prime\dagger}\hat \Delta'_{y}+\text{H.c.}$ Only half of the low-energy electrons participate in this interaction. We gap out the others with $H_m = \hat m_y +\text{H.c.}$, where 
$\hat m_{y} = \psi_{y-1,\uparrow,L}^\dagger \psi_{y,\uparrow,R}$ for even $y$ and $\hat m_{y} =\psi_{y-1,\downarrow,R}^\dagger\psi_{y,\downarrow,L}$ for odd $y$.\footnote{With unbroken $x$-translation symmetry, such terms require a staggered magnetic field and spin-orbit coupling.}
The ground state of $H_{\text{SC}'}=H_{\Delta'}+H_m$ features a spin gap and spontaneously breaks charge conservation. It also exhibits accidental edge states that depend on the termination. These do not play an important role here and are addressed below. 

Both terms in $H_{\text{SC}'}$ involve electrons hopping across wires and are suppressed once charges localize. Still, terms generated from these two interactions may persist across the Mott transition. Consider specifically $H^\text{spin}_{\text{SC}'} \equiv H_{\Delta'}|_{g \rightarrow \hat g} $ with $\hat g_{y}=\hat m_{y+1} \hat m_{y}$. Inside the superconductor, $\langle \hat g_{y}\rangle$ is of order unity, and $H_{\Delta'},H^\text{spin}_{\text{SC}'}$ are interchangeable. Crucially, the interactions in the latter, $e^{i4 \tilde \theta^s_{2y+1}}\equiv\hat\Delta^{\prime\dagger}_{2y+2}\hat\Delta^{\prime}_{2y+1}\hat g_{2y+1}$ and $e^{i2 \tilde \varphi^s_{2y}}\equiv\hat\Delta^{\prime\dagger}_{2y+1}\hat\Delta^{\prime}_{2y}\hat g_{2y}$, do not involve charge transfer between wires. Their phases, expressed through operators that satisfy canonical commutations with $\varphi^s_{2y+1}$ and $\theta^s_{2y}$, thus remain pinned across the Mott transition. All the charge transfer is contained in $H_m$, which takes on the role of $H_{\text{SC}'}^{\text{charge}}$.

In the Mott insulator, $H^\text{spin}_{\text{SC}'}$ realizes precisely the $\mathbb{Z}_2$ QSL described in Ref.~\onlinecite{Leviatan2020}. The QPT is described by $H_{\text{SC}'\text{--}\mathbb{Z}_2}\equiv \langle H^{\text{charge}}_{\text{SC}'}+H_{\text{Mott}}\rangle_{H^\text{spin}_{\text{SC}'}}$. We find
\begin{align}
\label{eq.O3}
 H_{\text{SC}'\text{--}\mathbb{Z}_2} = \cos[\tfrac{1}{2}(\tilde\varphi^c_{y+1}-\tilde\varphi^c_{y})]+ u\cos[4\theta^c_y]~,
\end{align}
with dressed charge operators $\tilde \varphi^c_{y}$ that are canonical conjugates to $\theta^c_y$ and avoid competition with $H^\text{spin}_{\text{SC}'}$. The same Hamiltonian describes the Mott transition of bosons $e^{i \tilde \varphi^c/2}$ at integer filling, which is in the universality class of the 3D XY model. In the present case, there is no local boson with unit charge, and the transition is refined to the XY$^*$ type.\cite{Senthil2000} The slowest fluctuating observable is the Cooper pair with anomalous exponent $\eta \approx 1.47$.\cite{Grover2010,Isakov2012} It is encoded as $e^{i\tilde\varphi^c_{y}} = \hat\Delta'_y \hat m_y e^{i2\tilde\theta^s_{y}}$ for odd $y$ and $e^{i\tilde\varphi^c_{y}} = \hat\Delta'_y \hat m_y e^{i\tilde\varphi^s_{y}}$ for even $y$.

To complete the analysis of the Mott transition, we trace the evolution of individual electrons into spinons. Consider the operator $\mathcal{O}_{2y'+1,2y} = \psi^\dagger_{2y'+1,\up,R}\psi_{2y,\up,L}$. We obtain $\mathcal{O}^\text{spin}$ by dressing it with the unique product of $\hat m$ that compensates for all inter-wire charge transfer. Inside the superconductor SC$'$, the bare and dressed operators are interchangeable. By construction, the latter evolves smoothly across the QPT. Finally, $\langle \mathcal{O}^\text{spin}\rangle_{H_\text{Mott}}$ yields the creation operator for a spinon particle-hole pair (see also App.~\ref{app.B1}).\cite{Leviatan2020} \\

\textit{Confinement transition---}The QPT between the $\mathbb{Z}_2$ QSL and the VBS is described by $H_{\mathbb{Z}_2\text{--}\text{VBS}}\equiv \langle \lambda H^\text{spin}_{\text{SC}'}+H^\text{spin}_{\text{SC}}+\delta H_{\text{VBS}}^\text{spin}\rangle_{H_
\text{Mott}}$. Notice that $\tilde\theta^s$ are pinned in both the VBS and QSL. Replacing them with $c$-numbers, we find
\begin{align}
\label{eq.XY_Clock}
 H_{\mathbb{Z}_2\text{--VBS}}= &u'\cos[\theta^s_{2y+2}-\theta^s_{2y}] + \lambda \cos[ 2 \tilde\varphi^s_{2y}] + \cos[4\theta^s_{2y}]~. 
\end{align}
The first two terms are equivalent to Eq.~\eqref{eq.O3}. The final term introduces a fourfold anisotropy, which is (dangerously) irrelevant at the 3D XY transition.\cite{Jose1977} Breaking wire-translation symmetry explicitly permits a strongly-relevant twofold anisotropy. Then, the confinement transition occurs independently of spatial symmetries and is described by a (dual) Ising model. 

The models above realize all the phases in Fig.~\ref{fig:phase_diagram}. To complete the phase diagram, we show that SC and SC$'$ are in the same phase. Recall that the latter exhibits accidental edge modes. Specifically, when terminating on an odd wire, the electron modes $\psi_{R,\downarrow}, \psi_{L,\uparrow}$ there are decoupled. These modes become gapped when coupled to a nearby region described by $H_\text{SC}$, suggesting no phase transition occurs. To verify that SC and SC$'$ are smoothly connected, we use that $\hat \Delta$ and $\hat \Delta '$ have non-zero expectation values in their respective superconductors. We thus describe SC and SC$'$ by free fermion models $H^\text{MF}_\text{SC} = \hat \Delta + \text{H.c.}$ and $H^\text{MF}_{\text{SC}'} = H_m +(\hat \Delta' + \text{H.c.})$ and find no gap-closure when tuning between them (see App.~\ref{app.A} for details).\\

\textit{Topological superconductor and non-Abelian QSL}---The simplest topological superconductor is comprised of spinless fermions.\cite{Read2000} To realize this phase, we thus begin by trivially gapping the $\downarrow$ electrons using $H_m$ and $H_\downarrow= \psi_{2y+1,\downarrow,R}^\dagger \psi_{2y,\downarrow,L} +\text{H.c.}$ The remaining electrons are effectively spinless. A topological superconductor can be obtained from them by diligently constructing inter-wire interactions such that a chiral Majorana fermion at the boundary remains uncoupled.\cite{TeoKane2014} Alternatively, one may use the fact that topological superconductor arises upon inducing pairing to Dirac electrons.\cite{Fu2008} A single Dirac cone for part of the $\uparrow$ electrons is realized by
\begin{align}
 H_{\uparrow,\text{Dirac}} = \psi_{4y-2,\uparrow,L}^\dagger[ \psi_{4y+1,\uparrow,R}- \psi_{4y-3,\uparrow,R}] + \text{H.c.}
\end{align}
To induce pairing without explicitly violating charge conservation, we first let the remaining electrons form a trivial (strongly paired) superconductor. Specifically, we take $H_{\uparrow,\text{SP}}= \hat\Delta^{\dagger}_{\uparrow,4y+3}\hat \Delta_{\uparrow,4y-1}+\text{H.c.}$, with the Cooper pair operator $\hat \Delta_{\uparrow,y} = \psi_{y+1,\uparrow,L} \psi_{y,\uparrow,R}$. Finally, the `proximity' coupling $H_{\uparrow,\Delta}= g\hat\Delta^{\dagger}_{\uparrow,2y+1}\hat \Delta_{\uparrow,2y-1}+\text{H.c.}$ 
yields a topological superconductor with a single chiral Majorana fermion at the edge. 

All terms in this model involve charge transfer between wires and are suppressed upon undergoing the Mott transition. However, as before, there are vestigial terms. To obtain the `spin' part of all couplings described above, we multiply them by the unique product of $\hat m$ operators that compensates for all inter-wire charge transfer. By construction, the resulting terms do not compete with the opening of the Mott gap. 

To identify the insulating phase, we note that $H_\downarrow^\text{spin}, H^\text{spin}_{\uparrow,\text{SP}}$ do not face competition and pin three operators per four-wire unit cell. The final mode plays an entirely different role. To reveal it, we define neutral fermions $f_{\chi}=e^{i\tilde\phi^\chi}$, with 
\begin{align}
 \nonumber \tilde\phi^\chi_{4y} \equiv & \varphi^s_{4y+1}-\varphi^s_{4y}+\theta^c_{4y}-\theta^c_{4y-1}-\theta^s_{4y}-\theta^s_{4y-1} \\
 & +2
 \begin{cases}
  \theta^c_{4y+1} & \ \quad\chi=R~, \\
  -\theta^c_{4y+2}-\theta^s_{4y+2}-\theta^s_{4y+1} & \ \quad\chi=L~.
 \end{cases}
\end{align}
Notice that inter-wire hopping of these fermions is not a local process. Still, they constitute deconfined excitations on top of the topologically non-trivial background formed by the pinned operators. Their effective Hamiltonian is $H_f\equiv \langle H_{\up,\text{Dirac}}^\text{spin}+H_{\up,\Delta}^\text{spin}\rangle
_{H^\text{spin}_{\downarrow}+H^\text{spin}_{\uparrow,\text{SP}}}$. We find
\begin{align}
\label{eq.NAQSL}
H_f= f^\dagger_{4y,L}[ f_{4y+4,R}- f_{4y,R}] + g f^\dagger_{4y,R}f^\dagger_{4y,L}+\text{H.c.}~,
\end{align}
which describes a neutral version of the electronic model discussed above. In particular, the $f$-fermions form a topological superconductor with a single chiral Majorana fermion at each edge. Consequently, the physical spin system realizes a non-Abelian QSL. The Mott transition is described by Eq.~\eqref{eq.O3} with modified microscopic expressions for $\tilde \varphi^c$ (see App.~\ref{app.B2} for details). Still, the Cooper-pair operator $\hat\Delta_{4y-1}\hat{m}_{4y}$ exhibits critical correlations with anomalous exponent $\eta\approx 1.47$.

Before we conclude this example, we note that a lattice model realizing the electronic band structure, $H_\down+H_m+H_{\up,\text{Dirac}}$, is readily constructed by engineering a suitable flux background (see App.~\ref{app.C} for details). Adding the pairing terms described above and an on-site repulsion then results in a lattice model for this non-Abelian QSL. \\

\textit{Quantum Hall insulators and chiral QSLs---}We turn to the Mott transition out of topological insulators. Here, the appealing perspective of Cooper pairs evolving into dimers is not applicable. Still, the techniques introduced above apply, and their implementation is even more straightforward. As a prototypical example of quantum Hall insulators, we consider a bilayer Laughlin state of bosons at filling factor $\nu_\sigma =\frac{1}{2}$. A parent Hamiltonian for this phase is the sum of \cite{Kane2002,TeoKane2014}
\begin{align}
 H_{220}^\text{charge} = \cos[2 \Theta^c_y]\cos[2 \Theta^s_y], \quad H_{220}^\text{spin} = \cos [4 \Theta^s_y]~,
\end{align}
with canonical variables
\begin{subequations}
\begin{align}
 \Phi^{c/s}_{y}&\equiv\tfrac{1}{4}(\varphi^{c/s}_{y}+\varphi^{c/s}_{y-1})+\theta_{y}^{c/s}-\theta_{y-1}^{c/s}~, \\
 \Theta^{c/s}_{y}&\equiv\tfrac{1}{4}(\varphi^{c/s}_{y}-\varphi^{c/s}_{y-1})+\theta_{y}^{c/s}+\theta_{y-1}^{c/s}~.
\end{align}
\end{subequations}
The insulator described by $\langle H^\text{spin}_{220}\rangle_{H_\text{Mott}}$ is a variant of the Kalmeyer-Laughlin chiral QSL,\cite{Kalmeyer1989} discussed in wire models by Refs.~\onlinecite{Gorohovsky2015,Meng2015}.

The nature of the QPT becomes apparent upon introducing fermions $\Psi^{c}_{R(L)}\equiv e^{i (\Phi^c \pm \Theta^c)}$. We find that $H_\text{$220$--CSL}\equiv \langle H_{220}^\text{charge}+H_{\text{Mott}}\rangle_{H^\text{spin}_{220}}$ reads
\begin{align}
 H_\text{$220$--CSL}= \Psi^{c,\dagger}_{L,y}[u\Psi^{c}_{R,y+1}-\Psi^{c}_{R,y}]+ \text{H.c.}~,\label{CSLtransition}
\end{align}
which describes a single-Dirac-cone band structure with mass $|1-u|$. Short-range interactions between Dirac fermions are irrelevant in two dimensions. Consequently, the transition is described by a single Dirac cone of free fermions that are spinless but carry unit electric charge. Individual fermions are non-local, but Cooper pairs represent pairs of physical bosons with opposite spins. At the QPT, this charge-$2e$ operator exhibits power-law correlations with scaling dimension $2$. By contrast, single-particle excitations remain gapped across the transition.

Other topological phases of bosons or fermions with \hbox{$2\times 2$} $K$-matrices can be analyzed analogously. The resulting Mott insulators are chiral QSLs unless the Hall conductance of the parent topological phase vanishes. For an example of the latter, consider a quantum spin Hall state described by $H_\text{QSH} = \sum_\sigma\psi_{y,\sigma, R}^\dagger \psi_{y+\sigma,\sigma,L}+\text{H.c.}$ Upon undergoing a Mott transition, only the correlated process of electrons swapping wires survives. In the insulating phase, where $\theta^c$ are replaced by $c$-numbers, we find this term to be $H^\text{spin}_\text{QSH} = \cos [\varphi^s_{y+1}-\varphi^s_y]$. It describes an easy-plane antiferromagnet, whose ground state spontaneously breaks U(1) spin-rotation symmetry. The Mott transition is described by Eq.~\eqref{eq.O3} with $\tilde \theta^c,\tilde\varphi^c$ replaced by
\begin{subequations}
\begin{align}
 \Theta_y &\equiv \tfrac{1}{8}(2\theta_{y+1}^s +2\theta_{y}^s-\varphi_{y+1}^c+\varphi_{y}^c)~,\\
 \Phi_y &\equiv\varphi^s_{y+1}+\varphi^s_{y}-2\theta^c_{y+1} +2\theta^c_{y}~,
\end{align}
\end{subequations}
respectively, and $u\rightarrow u^{-1}$. Therefore, the transition is in the universality class of the 3D XY model, agreeing with previous findings for the Kane-Mele-Hubbard model.\cite{kmhhachel,kmhlee,kmhdong,kmhgriset,khmhohenadler1} We note that interchanging spin and charge modes yields a transition from the quantum spin Hall state to the superconductor of the first example. [Ref.~\onlinecite{grover2008} studied a related transition for SU(2) symmetric models, which results in a different universality.]\\

\textit{Parton approach---}An alternative route for describing Mott transitions is based on the parton mean-field approach.\cite{ruckenstein1986,georges2002} Specifically, a microscopic electron (boson) $\psi_\sigma$ is expressed as $\psi_\sigma = c f_\sigma $, where $f_\sigma$ are fermionic (bosonic) `spinons' and $c$ is a bosonic `chargon.' The latter is at unit filling and, within a mean-field treatment, may undergo a conventional boson-Mott transition without changing the phase of the former. Fluctuations of the mean-field parameters take the form of a compact U(1) gauge field that couples to chargons and spinons.

When $f_\sigma$ form a superconductor, this emergent photon acquires a Higgs mass and does not modify the critical behavior. When $f_\sigma$ form a chiral phase, the emergent photon is rendered massive by a Chern-Simons term. Here, the transition is modified, as we found in Eq.~\eqref{CSLtransition}. Mean-field states where $f_\sigma$ form non-chiral insulators are unstable to monopole proliferation and require a different analysis. By contrast, our approach treats all cases on equal footing and allows us to derive their critical theories. We demonstrated this for VBS and QSH, confining phases within the parton approach.

The superconductor---VBS transition described by Eq.~\eqref{eq.SC_VBS} does not permit a straightforward description in terms of $c$ and $f_\sigma$. Fortunately, an alternative is suggested by the wire model. This QPT is related to the antiferromagnet---VBS QPT \cite{Leviatan2020} by interchanging spin and charge variables. Consequently, we propose the decomposition $\psi_\up = s h_+, \psi_\down=s^\dagger h_-^\dagger$. Specifically, the bosonic spinon $s$ is gapped on either side of the QPT, while the fermionic chargons $h_{\pm}$ transition from a trivial to a QSH insulator.

\textit{Discussion---}We adapted the coupled-wire approach for describing weak Mott insulators and the nearby itinerant phases. We used concrete models to understand what conditions favor topologically ordered Mott insulators over trivial VBS phases. In the superconductor denoted by SC, neighboring wires interact only through the fluctuations of $\hat \Delta$ around its expectation value, i.e., via the Goldstone mode. In particular, inter-wire spin correlations are strictly zero. Consequently, an intra-wire VBS phase is its natural fate after undergoing a Mott transition. Deforming the superconductor away from this limit into the one denoted by SC$'$ creates more non-trivial spin correlations, permitting a $\mathbb{Z}_2$ QSL to form.

Beyond the Mott transition itself, the weak-Mott-insulator lens has been conceptually useful for understanding exotic insulators.
This perspective becomes practically useful for generating parent Hamiltonians of these phases. Firstly, conventional coupled-wire constructions for spin-chain arrays involve carefully tuned and seemingly unnatural many-spin interactions. By contrast, the ones derived here originate in electron models with local hopping and two-body interactions only. Secondly, the latters' simplicity may make them more appealing than effective spin models for numerical techniques like the density matrix renormalization group.

Finally, the coupled-wire framework readily captures intrinsically non-mean-field states and even certain gapless QSLs.\cite{Rodrigo2018} Examples of the former are charge-$4e$ superconductors.\cite{Kivelson1990,Wu2005} There is no conceptual difference compared to the charge-$2e$ superconducting case, and there are no surprises. For completeness, we include the analysis in Appendix \ref{app.D}. The insulating phase is a $\mathbb {Z}_4$ QSL, and the transition is governed by the 3D XY$^{**}$ universality class, i.e., the critical correlations of the charge-$4e$ order parameter are determined by the fourth power of the classical XY order parameter. Extensions to gapless states would be an exciting direction for future studies.

This work was partially supported by the DFG (CRC/Transregio 183) and the ISF (2572/21).

\bibliography{mott.bib}

\begin{widetext}
\appendix

\section{Explicit bosonized expressions for the superconductors and QSLs}
\label{app.B}
This appendix provides additional details and explicit expressions for the superconductors and QSLs discussed in the main text. The bosonization convention specified there implies that the annihilation operators for low-energy electrons near the right (R) or left (L) Fermi points are
\begin{align}
  \psi_{y,\sigma,R/L} \sim \exp[i\phi^{R/L}_{y,\sigma}] \sim \exp[\tfrac{i}{2}(\varphi^c_y+\sigma\varphi^s_y \pm 2\theta^c_y \pm \sigma 2\theta^s_y)]~.
\end{align}
The operators $\varphi^{c/s}$ and $\theta^{c/s}$ form canonical pairs, i.e., satisfy
\begin{align}
  [\partial_x\theta^{c/s}_y(x),\varphi^{c/s}_{y'}(x')] = i\pi \delta(x-x')\delta_{y,y'}~,
\end{align}
while spin and charge modes commute. It follows that $\cos[ 2 \varphi(x)]$ and $\cos [2n \theta(x')]$ with $n\geq 2$ commute for $|x-x'|$ larger than the short-distance cutoff but not at shorter scales. Consequently, the two cosines cannot be simultaneously minimized. To avoid confusion and explicit references to short-distance physics, we use the terminology `competing' or `non-competing' instead of `commuting' or `non-commuting' in the main text.

\subsection{Kinetic terms}
The kinetic part of all the coupled-wire models we discuss is expressible as
\begin{align}
  H_\text{kin} = \frac{v}{2\pi}\int_x \sum_{y} [\partial_x \theta^c_{y},\partial_x \varphi^c_{y},\partial_x \theta^s_{y},\partial_x \varphi^s_{y}]^T V_{y,y'}[\partial_x \theta^c_{y},\partial_x \varphi^c_{y},\partial_x \theta^s_{y},\partial_x \varphi^s_{y}]~.
\end{align}
The (local) interaction matrix $V$ can always be taken to favor a particular set of inter-wire interactions, i.e., cosines.\cite{Kane2002} Once a particular set of cosines achieves dominance, $H_\text{kin}$ is strongly renormalized, making the specific form of the original $V$ of little consequence. Within fully gapped phases, all terms in $H_{\text{kin}}$ are strongly irrelevant in the renormalization group sense. At phase transitions, they contain kinetic terms for the critical fluctuations and additional short-range interactions between them. The latter do not affect the universality of the transition. For these reasons, we typically suppress the kinetic terms.

\subsection{BCS Superconductor and $\mathbb{Z}_2$ QSL}
\label{app.B1}
The bosonized form of the coupled-wire model realizing the superconducting phase with Cooper-pair operator
\begin{align}
\label{eq.Deltap}
  \hat{\Delta}'_{y} = \begin{cases}
    \psi_{y,\uparrow,R} \psi_{y-1,\downarrow,L} & y\text{ odd}~,\\
    \psi_{y,\uparrow,L} \psi_{y-1,\downarrow,R} & y\text{ even}~,
  \end{cases}
\end{align}
is given by $H_{\text{SC}'} = H_{\Delta'}+H_m$ with
\begin{align}
  H_{\Delta'} =& \int_x \sum_y g_y \cos[\tfrac{1}{2}(\varphi^c_{y+1}-\varphi^c_{y-1}+\varphi^s_{y+1}-2\varphi^s_{y}+\varphi^s_{y-1})+(-1)^y(\theta^c_{y+1}-\theta^c_{y-1}+\theta^s_{y+1}+2\theta^s_{y}+\theta^s_{y-1})]~, \\
  H_m =& \int_x \sum_y \cos[\tfrac{1}{2}(\varphi^c_{y+1}-\varphi^c_{y})-\tfrac{(-1)^y}{2}(\varphi^s_{y+1}+\varphi^s_{y})-(-1)^y(\theta^c_{y+1}+\theta^c_y)+\theta^s_{y+1}+\theta^s_y]~.
\end{align}
The `charge' and `spin' parts of $H_{\text{SC}'}$ take the form 
\begin{align}
  H^{\text{charge}}_{\text{SC}'} =& \int_x \sum_y (\cos[\tfrac{1}{2}(\tilde\varphi^c_{2y+1}-\tilde\varphi^c_{2y}+\tilde\varphi^s_{2y})+\tilde\theta^s_{2y+1}]+\cos[\tfrac{1}{2}(\tilde\varphi^c_{2y}-\tilde\varphi^c_{2y-1}+\tilde\varphi^s_{2y})+\tilde\theta^s_{2y-1}])~, \\
  H^{\text{spin}}_{\text{SC}'} =& \int_x \sum_y (\cos[2\tilde\varphi^s_{2y}]+\cos[4\tilde\theta^s_{2y+1}])~,
\end{align}
when written in terms of a new set of local operators, defined as 
\begin{subequations}
\begin{align}
  \tilde{\varphi}^s_{2y} \equiv& \varphi^s_{2y} -\tfrac{1}{2}(\varphi^s_{2y+1}+\varphi^s_{2y-1})-\theta^c_{2y+1}+\theta^c_{2y-1}~,\\
  \tilde{\theta}^s_{2y+1} \equiv& \theta^s_{2y+1} + \tfrac{1}{2}(\theta^s_{2y+2}+\theta^s_{2y}+\theta^c_{2y+2}-\theta^c_{2y})~,\\
  \tilde \varphi^c_{2y} \equiv& \varphi^c_{2y} +\tfrac{1}{2}(\varphi^s_{2y+1}-\varphi^s_{2y-1}) +\theta^c_{2y+1}+\theta^c_{2y-1}~,\\
  \tilde \varphi^c_{2y+1} \equiv& \varphi^c_{2y+1} -\theta^s_{2y+2} + \theta^s_{2y}-\theta^c_{2y+2} -\theta^c_{2y}~.
\end{align}
\end{subequations}
These are canonically conjugate to the original $\theta^s_{2y},\varphi^s_{2y+1},\theta^c_{2y},$ and $\theta^c_{2y+1}$, respectively. Crucially, since the original operators $\varphi^{c/s},\theta^{c/s}$ are locally expressible usign the new ones, $H_{\text{kin}}$ remains local, and its precise form is unimportant.

The operator $\mathcal{O}_{2y'+1,2y}=\psi^\dagger_{2y'+1,\up,R}\psi_{2y,\up,L}$ has the bosonized expression
\begin{align}
  \mathcal{O}_{2y'+1,2y} = \exp[\tfrac{i}{2}(\varphi^c_{2y'+1}-\varphi^c_{2y} + \varphi^s_{2y'+1}-\varphi^s_{2y})+i(\theta^c_{2y'+1}+\theta^c_{2y}+\theta^s_{2y'+1}+\theta^s_{2y})]~.
\end{align}
Similarly to obtaining $H^\text{spin}$, we construct an operator $\mathcal{O}_{2y'+1,2y}^\text{spin}$ that evolves smoothly across the Mott transition by stripping off its charge, i.e., $\mathcal{O}_{2y'+1,2y}^\text{spin}\equiv \mathcal{O}_{2y'+1,2y} \prod_{j=2y+1}^{2y'+1} \hat m_j $. Deep inside the superconductor SC$'$, where $\hat m_j$ have expectation values of order unity, $\mathcal{O}_{2y'+1,2y}$ and $\mathcal{O}^\text{spin}_{2y'+1,2y}$ are equivalent. The latter has the explicit form
\begin{align}
  \mathcal{O}_{2y'+1,2y}^\text{spin} =e^{ i (2\theta^c_{2y'+1}+2\theta^c_{2y})} e^{-i\sum_{j=2y}^{2y'+1}(-1)^j\varphi^s_j } 
  e^{-i \sum_{j=2y+1}^{2y'} 2\theta^s_j } 
  ~.
\end{align}
Projection of $\tilde{\mathcal{O}}_{2y'+1,2y}$ onto the ground state of $H_\text{Mott}$ amounts to replacing the first factor by a $c$-number. The remaining two factors are precisely the operators constructed in Ref.~\onlinecite{Leviatan2020} for creating the bosonic $e$ and $m$ excitations of the $\mathbb{Z}_2$ QSL. Together, they create a fermionic spinon.

\subsection{$p+ip$ superconductor and non-Abelian QSL}
\label{app.B2}
The bosonized form of the coupled-wire model realizing the $p+ip$ superconductor is given by $H_{p+ip} = H_\down + H_{\up,\text{Dirac}} + H_{\up,\Delta} + H_m$ with
\begin{align}
  \nonumber H_\down = \int_x \sum_y & \Big(\cos[\tfrac{1}{2}(\varphi^c_{2y+1}-\varphi^s_{2y+1}-\varphi^c_{2y}+\varphi^s_{2y})+(\theta^c_{2y+1}-\theta^s_{2y+1}+\theta^c_{2y}-\theta^s_{2y})] \\
  & \hspace{5pt} + \cos[\tfrac{1}{2}(\varphi^c_{2y+1}-\varphi^s_{2y+1}-\varphi^c_{2y}+\varphi^s_{2y})-(\theta^c_{2y+1}-\theta^s_{2y+1}+\theta^c_{2y}-\theta^s_{2y})]\Big)~, \\
  \nonumber H_{\up,\text{Dirac}} = \int_x \sum_y & \Big(w_{4y} \cos[\tfrac{1}{2}(\varphi^c_{4y-2}+\varphi^s_{4y-2}-\varphi^c_{4y-3}-\varphi^s_{4y-3})-\theta^c_{4y-2}-\theta^s_{4y-2}-\theta^c_{4y-3}-\theta^s_{4y-3}] \\
  & \hspace{5pt} + v_{4y}\cos[\tfrac{1}{2}(\varphi^c_{4y-2}+\varphi^s_{4y-2}-\varphi^c_{4y+1}-\varphi^s_{4y+1})-\theta^c_{4y-2}-\theta^s_{4y-2}-\theta^c_{4y+1}-\theta^s_{4y+1}]\Big)~, \\
  \nonumber H_{\up,\Delta} = \int_x \sum_y & g_{2y}\cos\Big[\tfrac{1}{2}(\varphi^c_{2y+2}+\varphi^s_{2y+2}+\varphi^c_{2y+1}+\varphi^s_{2y+1}-\varphi^c_{2y}-\varphi^s_{2y}-\varphi^c_{2y-1}-\varphi^s_{2y-1}) \\
  & \hspace{35pt} -\theta^c_{2y+2}-\theta^s_{2y+2}+\theta^c_{2y+1}+\theta^s_{2y+1}+\theta^c_{2y}+\theta^s_{2y}-\theta^c_{2y-1}-\theta^s_{2y-1}\Big]~, \\
  H_m = \int_x \sum_y & \cos[\tfrac{1}{2}(\varphi^c_{y+1}-\varphi^c_{y})-\tfrac{(-1)^y}{2}(\varphi^s_{y+1}+\varphi^s_{y})-(-1)^y(\theta^c_{y+1}+\theta^c_y)+\theta^s_{y+1}+\theta^s_y]]~.
\end{align}
The `charge' and `spin' parts of $H_{p+ip}$ take the form
\begin{align}
  \nonumber H^{\text{charge}}_{p+ip} = \int_x \sum_y \Big(&\cos[\tfrac{1}{2}(\tilde\varphi^c_{4y+3}-\tilde\varphi^c_{4y+2})]+\cos[\tfrac{1}{2}(\tilde\varphi^c_{4y+2}-\tilde\varphi^c_{4y+1})+\tilde\varPhi_{4y+2}]\\
  & + \cos[\tfrac{1}{2}(\tilde\varphi^c_{4y+1}-\tilde\varphi^c_{4y})]+\cos[\tfrac{1}{2}(\tilde\varphi^c_{4y}-\tilde\varphi^c_{4y-1})]\Big)~, \\
  \nonumber H^{\text{spin}}_{p+ip} = \int_x \sum_y \Big(&\cos[2\tilde\varPhi_{4y+2}] + \cos[2\tilde\theta^s_{4y+3}] + \cos[2\tilde\theta^s_{4y+1}] \\
  & + g_{4y}\cos[2\tilde\varphi_{4y}]+w_{4y+4}\cos[2\tilde\theta_{4y}]+v_{4y+4}\cos[\tilde\varphi_{4y+4}+\tilde\theta_{4y+4}-\tilde\varPhi_{4y+2}-\tilde\varphi_{4y}+\tilde\theta_{4y}]\Big)~.
\end{align}
The former is expressed using dressed operators
\begin{subequations}
\begin{align}
  \tilde \varphi^c_{4y} =& \varphi^c_{4y} + \varphi^s_{4y}+2\theta^c_{4y-1}+2\theta^s_{4y-1}~, \\
  \tilde \varphi^c_{4y+1} =& \varphi^c_{4y+1}-\varphi^s_{4y+1}+2\varphi^s_{4y}-2\theta^c_{4y+1}-2\theta^c_{4y}+2\theta^c_{4y-1}+2\theta^s_{4y+1}+2\theta^s_{4y}+2\theta^s_{4y-1}~, \\
  \tilde \varphi^c_{4y+2} =& \varphi^c_{4y+2} - \varphi^s_{4y+2} +2 \varphi^s_{4y+3} - 2\theta^c_{4y+4}+2\theta^c_{4y+3}+2\theta^c_{4y+2}-2\theta^s_{4y+4}-2\theta^s_{4y+3}-2\theta^s_{4y+2}~, \\
  \tilde \varphi^c_{4y+3} =& \varphi^c_{4y+3}+\varphi^s_{4y+3}-2\theta^c_{4y+4}-2\theta^s_{4y+4}~. 
\end{align}
\end{subequations}
which are conjugate to the original $\theta^c$. For the spin part, we use the conjugate pairs
\begin{subequations}
\begin{align}
  \nonumber \tilde \varPhi_{4y+2} =& -\varphi^s_{4y+3}+\varphi^s_{4y+2}-\varphi^s_{4y+1}+\varphi^s_{4y}+\theta^c_{4y+4}-\theta^c_{4y+3}-\theta^c_{4y}+\theta^c_{4y-1}\\
  &+\theta^s_{4y+4}+\theta^s_{4y+3}+2\theta^s_{4y+2}+2\theta^s_{4y+1}+\theta^s_{4y}+\theta^s_{4y-1}~, \\
  \tilde \varTheta_{4y+2} =& \theta^s_{4y+2}+\theta^c_{4y+2}~, \\
  \tilde \varphi^s_{4y+3} =& \varphi^s_{4y+3}+\theta^c_{4y+6}+\theta^c_{4y+5}+2\theta^c_{4y+2}+\theta^s_{4y+6}+\theta^s_{4y+5}~, \\
  \tilde \theta^s_{4y+3} =& \theta^s_{4y+3}+\theta^s_{4y+2}-\theta^c_{4y+3}-\theta^c_{4y+2}~, \\
  \tilde \varphi^s_{4y+1} =& \varphi^s_{4y}+\theta^c_{4y+2}-\theta^c_{4y+1}+2\theta^c_{4y-1}+2\theta^c_{4y-2}+\theta^s_{4y+2}+\theta^2_{4y+1}~, \\
  \tilde \theta^s_{4y+1} =& \theta^s_{4y+1}+\theta^s_{4y}-\theta^c_{4y+1}-\theta^c_{4y}~, \\
  \tilde \varphi_{4y} =& \varphi^s_{4y+1}-\varphi^s_{4y}-\theta^c_{4y+2}+\theta^c_{4y+1}+\theta^c_{4y}-\theta^c_{4y-1}-\theta^s_{4y+2}-\theta^s_{4y+1}-\theta^s_{4y}-\theta^s_{4y-1}~, \\
  \tilde \theta_{4y} =& \theta^c_{4y+2}+\theta^c_{4y+1}+\theta^s_{4y+2}+\theta^s_{4y+1}~. 
\end{align}
\end{subequations}
As in the $\mathbb{Z}_2$ case, the kinetic Hamiltonian remains local with new variables, and its precise form is thus unimportant. The first line of $H^{\text{spin}}_{p+ip}$ contains only cosines with non-competing arguments. We thus refer to it as the Abelian part in the text. Upon replacing $\tilde\varPhi$ with a $c$-number, the second line becomes the bosonized version of $H_f$ presented in the main text. The fermions $f\sim e^{i\tilde\phi^{R/L}}$ are constructed from the variables introduced above as $\tilde\phi^{R/L}_{4y}=\tilde\varphi_{4y}\pm\tilde\theta_{4y}$.

\section{Adiabatically connecting the superconductors SC and SC'}
\label{app.A}
Here, we show that the two coupled-wire models
\begin{align}
  H^{\text{MF}}_{\text{SC}} = \int_x \sum_y &\Delta [\psi_{y,\up,R}\psi_{y,\down,L}-\psi_{y,\down,R}\psi_{y,\up,L}+\text{H.c.}]~,\\
  H^{\text{MF}}_{\text{SC}'} =\int_x \sum_y \Big(& \Delta'[\psi_{2y+1,\up,R}\psi_{2y,\down,L}+\psi_{2y,\down,L}\psi_{2y-1,\up,R}] + m[\psi^\dagger_{2y+1,\down,L}\psi_{2y,\down,R}+\psi^\dagger_{2y,\up,R}\psi_{2y-1,\up,L} ]+\text{H.c.}\Big)~,
\end{align}
describe the same phase. We supplement these pairing and inter-wire terms by the kinetic energy $H_0 \sim \sum_{\sigma} i[\psi^\dagger_{y,\sigma,R} \partial_x\psi_{y,\sigma,R}-\psi^\dagger_{y,\sigma,L} \partial_x\psi_{y,\sigma,L}]$ and compute the spectrum of $H = H_0 + H_{\text{SC}}^{\text{MF}}+H_{\text{SC}'}^{\text{MF}}$. 
We label the two wires in each unit cell $j=(2y,2y+1)$ by $\psi_{2y(+1)}=\psi_{j,e(o)}$. Transforming to $k$-space, we find $H = \int d^2k \Psi^\dagger(\vec{k}) h(\vec{k}) \Psi(\vec{k})$ with
\begin{align}
  \Psi(\vec{k}) &= \begin{pmatrix} \psi_{\vec{k},o,\up,L} & \psi_{\vec{k},e,\up,R} & \psi_{\vec{k},o,\up,R} & \psi_{\vec{k},e,\up,L} & \psi^\dagger_{-\vec{k},o,\down,R} & \psi^\dagger_{-\vec{k},e,\down,L} & \psi^\dagger_{-\vec{k},o,\down,L} & \psi^\dagger_{-\vec{k},e,\down,R} \end{pmatrix}^T~,\\
  h(\vec{k}) &= \begin{pmatrix} 
    -k_x & m e^{ik_y} & 0 & 0 & -\Delta & 0 & 0 & 0 \\
    me^{-ik_y} & k_x & 0 & 0 & 0 & -\Delta & 0 & 0 \\
    0 & 0 & k_x & 0 & 0 & -\Delta' & -\Delta & 0 \\
    0 & 0 & 0 & -k_x & -\Delta'e^{-ik_y} & 0 & 0 & -\Delta \\
    -\Delta & 0 & 0 & -\Delta'e^{ik_y} & k_x & 0 & 0 & 0 \\
    0 & -\Delta & -\Delta' & 0 & 0 & -k_x & 0 & 0 \\
    0 & 0 & -\Delta & 0 & 0 & 0 & -k_x & -m \\
    0 & 0 & 0 & -\Delta & 0 & 0 & -m & k_x
    \end{pmatrix}~.
\end{align}
The eigenvalues of $h$ are readily obtained. Rescaling $\Delta\rightarrow\sqrt{2}\Delta$ and setting, for simplicity, $m=\Delta'$, we find $\epsilon_\pm^\pm(\vec{k}) = \pm \sqrt{k_x^2+\Delta^2+(\Delta\pm\Delta^{\prime })^2}$, which are two-fold degenerate. Crucially, the gap does not close when tuning from $\Delta\neq 0$ and $\Delta'=m=0$ to $\Delta =0$ and $\Delta',m' \neq 0$. Consequently, SC and SC$'$ are representatives of the same phase. 

\section{Electronic lattice model realizing a non-Abelian spin liquid}
\label{app.C}
A lattice model that produces, at low energy, the coupled-wire model for the $p+ip$ superconductor is readily constructed. First, the LL array arises from nearest-neighbor hopping in the $x$-direction, i.e.,
\begin{align}
  H_t = \sum_{x,y}\sum_\sigma (-t\psi^\dagger_{x+1,y,\sigma}\psi_{x,y,\sigma}+\text{H.c.})~,
\end{align}
with $t>0$. Second, the desired band structure for the electrons is obtained by including hopping in the $y$-direction. A diligent choice of background fluxes frustrates certain hoppings and eliminates them from the low-energy description. Specifically, we take the inter-wire hopping
\begin{subequations}
\begin{align}
  H_{t'} &= \sum_{x,y} [t'(-1)^x\psi^\dagger_{x,2y+1,\down}\psi_{x,2y,\down}+\text{H.c.}]~,\\
  H_{t''} &= \sum_{x,y} [ t''(-1)^x(\psi^\dagger_{x,2y,\up}\psi_{x,2y-1,\up}+\tfrac{i}{2}\psi^\dagger_{x+1,2y,\up}\psi_{x,2y-1,\up}+\tfrac{i}{2}\psi^\dagger_{x,2y,\up}\psi_{x+1,2y-1,\up}) + \text{H.c.}]~, \\
  H_u &= \sum_{x,y}[u(-1)^x(\psi^\dagger_{x,4y-2,\up}\psi_{x,4y+1,\up}+\tfrac{i}{2}\psi^\dagger_{x+1,4y-2,\up}\psi_{x,4y+1,\up}+\tfrac{i}{2}\psi^\dagger_{x,4y-2,\up}\psi_{x+1,4y+1,\up})+ \text{H.c.}]~,\\
  H_v &= \sum_{x,y}[v(-1)^x(\psi^\dagger_{x,4y-2,\up}\psi_{x,4y-3,\up}+\tfrac{i}{2}\psi^\dagger_{x+1,4y-2,\up}\psi_{x,4y-3,\up}+\tfrac{i}{2}\psi^\dagger_{x,4y-2,\up}\psi_{x+1,4y-3,\up})+ \text{H.c.}]~,
\end{align}
\end{subequations}
with coefficients $t',t'',u,v$. To realize the wire limit, these must be smaller than $t$. $H_{t'}$ provides $H_\down$ and $H_m$ on odd wires. These are responsible for the gapping of the $\down$ electrons. Similarly, $H_{t''}$ implements $H_m$ on the even wires, gapping some of the $\up$ electrons. Finally, $H_v + H_u$ account for $H_{\up,\text{Dirac}}$ that produces the Dirac band-structure for the remaining $\up$ electrons.

The final contribution to $H_{p+ip}$, i.e., the pairing term $H_{\up,\Delta}$, is realized by
\begin{align}
  H_g = \sum_{x,y} (g\hat\Delta^{\dagger}_{x,2y+1,\uparrow}\hat \Delta_{x,2y-1,\uparrow}+ \text{H.c.})~,
\end{align}
with $\hat \Delta_{x,2y-1,\uparrow} = \psi_{x,2y,\uparrow} \psi_{x,2y-1,\uparrow} + \tfrac{i}{2} \psi_{x+1,2y,\uparrow} \psi_{x,2y-1,\uparrow}+\tfrac{i}{2}\psi_{x,2y,\uparrow} \psi_{x+1,2y-1,\uparrow}$. Importantly, $g$ must be larger than $|v-u|$, which may be taken to zero for simplicity. The specific form of $\hat \Delta_{x,2y-1,\uparrow}$ is required to generate exactly $H_{\up,\Delta}$, which appears in the main text. However, replacing it with $\hat \Delta'_{x,2y-1,\uparrow} = \psi_{x,2y,\uparrow} \psi_{x,2y-1,\uparrow}$ is sufficient to realize the same $p+ip$ superconductor. In the wire limit, $\hat \Delta'_{x,2y-1,\uparrow}$ contains the desired $\psi_{2y,\uparrow,L} \psi_{x,2y-1,\uparrow,R}$ as well as its counterpart with opposite chirality $R\leftrightarrow L$. The latter (but not the former) renormalizes to zero due to its competition with $H_{t'}+H_{t''}$. Finally, to obtain a lattice model for the non-Abelian spin liquid presented in the main text, one needs only further include the on-site repulsion $H_{\text{Mott}}\sim \psi_{x,y,\up}^\dagger\psi_{x,y,\down}^\dagger\psi_{x,y,\up}\psi_{x,y,\down}+\text{H.c.}$, with a large coefficient. 

\section{Charge-$4e$ superconductor and $\mathbb{Z}_4$ QSL}
\label{app.D}
We now demonstrate our approach for an intrinsically non-mean-field example, namely a charge-$4e$ superconductor. Its order parameter is the ``Cooper-quartet'' operator
\begin{align}
  \hat\Delta^{4e}_{2y}= \begin{cases}
    \hat\Delta'_{2y}\hat\Delta'_{2y-2} & $y$\text{ even}~, \\
    \hat\Delta'_{2y+1}\hat\Delta'_{2y-1} & $y$\text{ odd}~,
  \end{cases}
\end{align}
with $\hat\Delta'$ of Eq.~\eqref{eq.Deltap}. To realize this phase, we begin with a hopping term for $\Delta^{4e}$, i.e., $H_{\Delta^{4e}}= \int_x \sum_y \hat g_{2y+1}\Delta^{4e\dagger}_{2y+2}\hat\Delta^{4e}_{2y}+\text{H.c.}$ This interaction does not gap out all modes. Firstly, $\hat\Delta'$ does not involve the fermion modes that comprise $\hat m_y$, and we thus add $H_m$. Secondly, $H_{\Delta^{4e}}$ and $H_m$ together still leave three unpinned modes in every four-wire unit cell. To obtain a fully gapped charge-$4e$ superconductor, we must gap out these remaining modes without generating a non-zero expectation value for any local charge-$2e$ operator. We thus include
\begin{align}
  H' =& \int_x \sum_y [\psi^\dagger_{4y+3,\up,R} \psi_{4y+2,\up,L}(h_{4y+2} \psi^\dagger_{4y,\down,L} \psi_{4y+3,\down,R} + h_{4y+3}\psi^\dagger_{4y+4,\up,L} \psi_{4y+1,\up,R}) + \text{H.c.}]~,
\end{align}
i.e., the full charge-$4e$ superconductor model is $H_{4e\text{-SC}} = H_{\Delta^{4e}} + H_m + H'$. All terms in $H_{4e\text{-SC}}$ include the inter-wire transfer of charges and thus compete with $H_{\text{Mott}}$. To obtain $H_{4e\text{-SC}}^\text{spin}$, we thus promote 
\begin{align}
  g_{4y+3} \rightarrow& \hat g_{4y+3} = \hat m_{4y+4}\hat m_{4y+3}\hat m_{4y+2}\hat m_{4y+1}~,\\
  g_{4y+1} \rightarrow& \hat g_{4y+1} = \hat m_{4y+3} \hat m^2_{4y+2} \hat m^3_{4y+1} \hat m^3_{4y} \hat m^2_{4y-1} \hat m_{4y-2}~,\\
  h_{4y+2} \rightarrow& \hat h_{4y+2} = \hat m_{4y+2}^\dagger \hat m_{4y+1}^\dagger~,\\
  h_{4y+3} \rightarrow& \hat h_{4y+3} = \hat m_{4y+4} \hat m^2_{4y+3}\hat m_{4y+2}~.
\end{align}
Explicitly, $H_{4e\text{-SC}}^\text{spin}$ reads
\begin{align}
  \nonumber H^{\text{spin}}_{4e\text{-SC}} =& \int_x \sum_y (\cos[2\varPhi^s_{4y+2}] + \cos[4\tilde\theta^s_{4y+2}] + \cos[2\overline\varphi^s_{4y+2}-2\varPhi^s_{4y+2}+4\tilde\theta^s_{4y+2}] + \cos[2\overline{\varphi}^s_{4y+2}+4\tilde\varphi^2_{4y}+2\overline{\varphi}^s_{4y-2}])~,
\end{align}
which is equivalent to
\begin{align}
  \nonumber H^{\text{spin}}_{4e\text{-SC}}\rightarrow & \int_x \sum_y (\cos[2\varPhi^s_{4y+2}] + \cos[4\tilde\theta^s_{4y+2}] + \cos[2\overline\varphi^s_{4y+2}] + \cos[4\tilde\varphi^s_{4y}])~.
\end{align}
The canonical operator-pairs introduced here are
\begin{subequations}
\begin{align}
  \tilde\varphi^s_{4y} =& \varphi^s_{4y} -\tfrac{1}{2}(\varphi^s_{4y+1}+\varphi^s_{4y-1}-\theta^s_{4y+2}-2\theta^s_{4y+1}-2\theta^s_{4y}-2\theta^s_{4y-1}-\theta^s_{4y-2}-\theta^c_{4y+2}+2\theta^c_{4y+1}-2\theta^c_{4y-1}+\theta^c_{4y-2})~, \\
  \tilde\theta^s_{4y} =& \theta^s_{4y} ~, \\
  \tilde\varphi^s_{4y+2} =& \varphi^s_{4y+2} - \theta^s_{4y+4}-\theta^s_{4y+3}+\theta^s_{4y+1}~, \\
  \tilde\theta^s_{4y+2} =& \theta^s_{4y+2} + \tfrac{1}{2}(\theta^c_{4y+4}+\theta^s_{4y+4}+2\theta^s_{4y+3}+2\theta^s_{4y+1}+\theta^s_{4y}-\theta^c_{4y})~, \\
  \overline\varphi^s_{4y+2} =& -\tfrac{1}{2}(\varphi^s_{4y+3}-2\varphi^s_{4y+2}+\varphi^s_{4y+1}+2\theta^c_{4y+3}-2\theta^c_{4y+1})~, \\
  \overline\theta^s_{4y+2} =& -\tfrac{1}{2}(\theta^s_{4y+4}+2\theta^s_{4y+3}+2\theta^s_{4y+1}+\theta^s_{4y})~, \\
  \varPhi^s_{4y+2} =& \tfrac{1}{2}(\varphi^s_{4y+3}-\varphi^s_{4y+1}+2\theta^s_{4y+3}+2\theta^s_{4y+2}+2\theta^s_{4y+1}+2\theta^s_{4y}+2\theta^c_{4y+2}-2\theta^c_{4y})~, \\
  \varTheta^s_{4y+2} =& \tfrac{1}{2}(\theta^s_{4y+4}+2\theta^s_{4y+3}-2\theta^s_{4y+1}-\theta^s_{4y})~. 
\end{align}
\end{subequations}
These commute with dressed charge operators 
\begin{subequations}
\begin{align}
  \tilde\varphi^c_{4y} =& \varphi^c_{4y} +\tfrac{1}{2}(\varphi_{4y+2}^{s}-\varphi_{4y-2}^{s}+\theta_{4y+3}^{s}-\theta_{4y+1}^{s}+\theta_{4y-1}^{s}-\theta_{4y-3}^{s}-\theta_{4y+3}^{c}+3\theta_{4y+1}^{c}+3\theta_{4y-1}^{c}-\theta_{4y-3}^{c})~, \\
  \tilde\varphi^c_{4y+1} =& \varphi^c_{4y+1} +\tfrac{1}{2}(\theta_{4y+4}^{s}+2\theta_{4y+3}^{s}+2\theta_{4y+1}^{s}+3\theta_{4y}^{s}+\theta_{4y+4}^{c}-2\theta_{4y+2}^{c}-3\theta_{4y}^{c})~, \\
  \tilde\varphi^c_{4y+2} =& \varphi^c_{4y+2} -\theta_{4y+3}^{s}+\theta_{4y+1}^{s}+\theta_{4y+3}^{c}+\theta_{4y+1}^{c}~, \\
  \tilde\varphi^c_{4y+3} =& \varphi^c_{4y+3} -\tfrac{1}{2}(3\theta_{4y+4}^{s}+2\theta_{4y+3}^{s}+2\theta_{4y+1}^{s}+\theta_{4y}^{s}+3\theta_{4y+4}^{c}+2\theta_{4y+2}^{c}-\theta_{4y}^{c})~,
\end{align}
\end{subequations}
which are conjugate to the original $\theta^c$. The `charge' Hamiltonian is given by \begin{align}
  \nonumber H^{\text{charge}}_{4e\text{-SC}} = \int_x \sum_y \Big(&\cos[\tfrac{1}{2}(\tilde\varphi^c_{4y+3}-\tilde\varphi^c_{4y+2}-\Phi_{4y+2}^{s}+3\tilde{\theta}_{4y+2}^{s}+\overline{\varphi}_{4y+2}^{s})]\\
  \nonumber &+ \cos[\tfrac{1}{2}(\tilde\varphi_{4y+2}^{c}-\tilde\varphi_{4y+1}^{c}+\Phi_{4y+2}^{s}+\tilde{\theta}_{4y+2}^{s}+\overline{\varphi}_{4y+2}^{s})]\\
  \nonumber &+\cos[\tfrac{1}{2}(\tilde\varphi_{4y+1}^{c}-\tilde\varphi_{4y}^{c}+\tfrac{1}{2}\Phi_{4y+2}^{s}-\tilde{\theta}_{4y+2}^{s}+\tfrac{1}{2}\overline{\varphi}_{4y+2}^{s}+\tilde{\varphi}_{4y}^{s}+\tfrac{1}{2}\Phi_{4y-2}^{s}-\tilde{\theta}_{4y-2}^{s}-\tfrac{1}{2}\overline{\varphi}_{4y-2}^{s})]\\
  &+\cos[\tfrac{1}{2}(\tilde\varphi_{4y}^{c}-\tilde\varphi_{4y-1}^{c}-\tfrac{1}{2}\Phi_{4y+2}^{s}-\tfrac{1}{2}\overline{\varphi}_{4y+2}^{s}+\tilde{\varphi}_{4y}^{s}+\tfrac{1}{2}\overline{\varphi}_{4y-2}^{s}-\tfrac{1}{2}\Phi_{4y-2}^{s})]~,
\end{align}
and the transition is therefore described by
\begin{align}
  H_{4e\text{-SC}-\mathbb{Z}_4} \equiv \langle H_{4e\text{-SC}}+H_{\text{Mott}} \rangle_{H^{\text{spin}}_{4e\text{-SC}}} = \int_x \sum_y (\cos[\tfrac{1}{2}(\tilde\varphi^c_{y+1}-\tilde\varphi^c_{y})]+ u \cos[4\theta^c_y])~.
\end{align}
This effective Hamiltonian puts the transition in the universality class of the XY model in three dimensions. Notice, however, that only the fourth power of the XY order-parameter $e^{i \tilde \varphi^c/2}$ is physical, analogous to the XY$^*$ transition where only its square is physical.

The spin model describing the Mott insulating phase is obtained by projecting $H^{\text{spin}}_{4e\text{-SC}}$ onto the ground state of $H_{\text{Mott}}$. In bosonized form, this amounts to replacing $\theta^c$ by a $c$-number in $H^{\text{spin}}_{4e\text{-SC}}$. In terms of the original variables, we find
\begin{align}
  \nonumber H_{\mathbb{Z}_4}\equiv\langle H^{\text{spin}}_{4e\text{-SC}} \rangle_{H_{\text{Mott}}} =& \int_x \sum_y ( \cos[\varphi^s_{4y+3}-2\varphi^s_{4y+2}+\varphi^s_{4y+1}] + \cos[2\theta^s_{4y+4}+4\theta^s_{4y+3}+4\theta^s_{4y+2}+4\theta^s_{4y+1}+2\theta^s_{4y}])\\
  \nonumber & + \int_x \sum_y \cos[ 2\varphi^s_{4y+1}-4\varphi^s_{4y}+2\varphi^s_{4y-1}-2\theta^s_{4y+2}-4\theta^s_{4y+1}-4\theta^s_{4y}-4\theta^s_{4y-1}-2\theta^s_{4y-2}] \\ 
  & + \int_x \sum_y \cos[\varphi^s_{4y+3}-\varphi^s_{4y+1}+2\theta^s_{4y+3}+2\theta^s_{4y+2}+2\theta^s_{4y+1}+2\theta^s_{4y}]~,
\end{align}
which realizes a $\mathbb{Z}_4$ QSL. Its topological excitations are created by
\begin{subequations}
\begin{align}
  \mathcal{O}_A\left[4y_2,4y_1\right]&= e^{i\sum\nolimits_{y=y_1}^{y_2-1}(\varphi^s_{4y+2}-\varphi^s_{4y+1})} = d^\dagger_{4y_2,A} s_A\left[4y_2,4y_1\right] d_{4y,A}~,\\
  \mathcal{O}_B\left[4y_2,4y_1\right] &= e^{i\sum\nolimits_{y=y_1+1}^{y_2}(\varphi^s_{4y+1}-\varphi^s_{4y})} = d^\dagger_{4y_2+1,B} s_B\left[4y_2,4y_1\right] d_{4y_1+1,B}~,
\end{align}
\end{subequations}
where $d_{4y,A} = e^{-\frac{i}{2}\theta^s_{4y}}$, $d_{4y+1,B}=e^{-\frac{i}{2}(\varphi^s_{4y+1}+\theta^s_{4y+2}+2\theta^s_{4y+1})}$, and $s_{A/B}$ are exponentials of pinned operators. Explicitly,
\begin{align}
  \nonumber s_A\left[4y_2,4y_1\right] = \exp\Bigg\{i\sum_{y=y_1}^{y_2-1} \Big(&\tfrac{1}{2}[\theta^s_{4y+4}+2\theta^s_{4y+3}+2\theta^s_{4y+2}+2\theta^s_{4y+1}+\theta^s_{4y}]+\tfrac{1}{2}[\varphi^s_{4y+3}-2\varphi^s_{4y+2}+\varphi^s_{4y+1}]\\
  &-\tfrac{1}{2}[\varphi^s_{4y+3}-\varphi^s_{4y+1}+2\theta^s_{4y+3}+2\theta^s_{4y+2}+2\theta^s_{4y+1}+2\theta^s_{4y}]\Big)\Bigg\}~,\\
  \nonumber s_B\left[4y_2,4y_1\right] = \exp\Bigg\{i\sum_{y=y_1}^{y_2-1} \Big(&\tfrac{1}{4}[2\varphi^s_{4y+5}-4\varphi^s_{4y+4}+2\varphi^s_{4y+3}-2\theta^s_{4y+6}-4\theta^s_{4y+5}-4\theta^s_{4y+4}-4\theta^s_{4y+3}-2\theta^s_{4y+2}]\\
  \nonumber &+\tfrac{1}{2}[2\theta^s_{4y+4}+4\theta^s_{4y+3}+4\theta^s_{4y+2}+4\theta^s_{4y+1}+2\theta^s_{4y}]\\
  &-\tfrac{1}{2}[\varphi^s_{4y+3}-\varphi^s_{4y+1}+2\theta^s_{4y+3}+2\theta^s_{4y+2}+2\theta^s_{4y+1}+2\theta^s_{4y}]\Big)\Bigg\}~.
\end{align}
In the ground state of $H_{\mathbb{Z}_4}$, these strings are $c$-numbers and thus not observable. When both $A$ and $B$ excitations are present, the braiding properties can be obtained from the commutator of $d_A$ with $s_B$ or $d_B$ with $d_A$.\cite{TeoKane2014} We find that a full braid between the two results in a phase of $\pi/2$, as expected for a $\mathbb{Z}_4$ QSL.

\end{widetext}

\end{document}